\documentclass[preprint,12pt]{elsarticle}

\usepackage{arydshln}
\usepackage{verbatimbox}

\usepackage{graphicx}
\usepackage{subfig}
\usepackage{url}
\usepackage{hyperref}

\usepackage{amsmath,amssymb,amsfonts}
\usepackage[linesnumbered,lined,boxed,commentsnumbered,ruled]{algorithm2e}
\usepackage{textcomp}
\usepackage{multirow}
\usepackage{multicol}
\usepackage{color}
\usepackage{natbib}

\usepackage{caption}
\usepackage{setspace}
\usepackage{comment}

\usepackage{threeparttable}
\usepackage{booktabs}
\usepackage{tikz}
\usetikzlibrary{arrows,positioning}
\usepackage{lipsum, adjustbox}

\begin{document}

\let\WriteBookmarks\relax
\def\floatpagepagefraction{1}
\def\textpagefraction{.001}

\title{Moving from Cross-Project Defect Prediction to Heterogeneous Defect Prediction: A Partial Replication Study}                      

\author{Hadi Jahanshahi} 
\ead{hadi.jahanshahi@ryerson.ca}
\author{Mucahit Cevik}
\author{Ay\c{s}e Ba\c{s}ar}


\address{Data Science Lab at Ryerson University, Toronto, ON M5B 1G3, Canada}


\begin{abstract}
Software defect prediction heavily relies on the metrics collected from software projects. Earlier studies often used machine learning techniques to build, validate, and improve bug prediction models using either a set of metrics collected within a project or across different projects. However, techniques applied and conclusions derived by those models are restricted by how identical those metrics are. Knowledge coming from those models will not be extensible to a target project if no sufficient overlapping metrics have been collected in the source projects.
To explore the feasibility of transferring knowledge across projects without common labeled metrics, we systematically integrated Heterogeneous Defect Prediction (HDP) by replicating and validating the obtained results. Our main goal is to extend prior research and explore the feasibility of HDP and finally to compare its performance with that of its predecessor, Cross-Project Defect Prediction.
We construct an HDP model on different publicly available datasets. Moreover, we propose a new ensemble voting approach in the HDP context to utilize the predictive power of multiple available datasets.
The result of our experiment is comparable to that of the original study. However, we also explored the feasibility of HDP in real cases. Our results shed light on the infeasibility of many cases for the HDP algorithm due to its sensitivity to the parameter selection.
In general, our analysis gives a deep insight into why and how to perform transfer learning from one domain to another, and in particular, provides a set of guidelines to help researchers and practitioners to disseminate knowledge to the defect prediction domain.
\end{abstract}


\maketitle

\section{INTRODUCTION}\label{sec:Intro}
Defect prediction is a powerful technique proposed to allocate limited testing resources efficiently and minimise the risk associated with incurring post-release defects. This technique helps software teams to prioritise potential defective modules (e.g., files or functions) in the software systems. In most cases, the metrics extracted from the historical defect dataset of a project are used to train a binary classifier to predict the defects of new software modules from the same project. This type of prediction is called Within-Project Defect Prediction (WPDP). However, this approach does not apply to the case where new software is launched, and there is no significant evidence about the characteristics of bugs in the system.  

In Cross-Project Defect Prediction (CPDP) \cite{Turhan-cross, Zimmermann-CPDP}, we may use another project, with the same metrics as those of the testing set, to build a model and create a promising prediction. CPDP aims to predict bugs in a new project that lacks historical defect data. This process constructs a model from a similar but not identical project. Nevertheless, in practice, collecting datasets with the same metrics is not feasible for all projects. To alleviate the limitations due to heterogeneous metric sets in CPDP, \citet{HDP} proposed Transfer Learning or Heterogeneous Defect Prediction (HDP). They presented an HDP method exploiting gain ratio feature selection and metrics matching based on the Kolmogorov-Smirnov test (KS-test) to choose the metrics with similar distribution in the training set. There are other variations of the HDP method. \citet{gong2019unsupervised} suggested an unsupervised deep domain adaption method to overcome the issues with heterogeneous metrics and unbalanced datasets. \citet{li2019heterogeneous} proposed a two-stage ensemble learning for HDP to alleviate the problem of linear inseparability and class imbalance.

We first partially replicate the work by \citet{HDP}, and then investigate new research questions on the possibility of transferring the lessons learned from traditional defect prediction to a new project with different features. To systematically explore the issue, we construct our study along with the following four research questions:

\begin{itemize}
    \item \textbf{RQ1}: How do models selected using domain-agnostic similarity perform in a cross-project context?

    \item \textbf{RQ2}: How do HDP methods predict the defect in the system compared to WPDP methods?

    \item \textbf{RQ3}: How do the ensembles of models built from several projects perform in an HDP context?

    \item \textbf{RQ4}: How feasible is HDP in terms of target prediction coverage? 
\end{itemize}

The remainder of the paper is organised as follows. Section~\ref{sec:paper2} presents a brief background of the original study. In Section~\ref{sec:Info_about_rep}, we provide an overview of the replication study. Afterward, In Section~\ref{sec:comparison}, the comparison of the result with the original study has been discussed. Moreover, we report the performance of the approach under different scenarios and investigate the coverage ability of HDP. Section~\ref{sec:validity} discusses the threats to the validity of the paper, and finally, Section~\ref{sec:Conclusion} concludes the paper.

\section{INFORMATION ABOUT THE ORIGINAL STUDY} \label{sec:paper2}
This paper reports on the replication and extension of a paper, established upon the proposed guidelines for experimental replications~\cite{replication}. 

\citet{HDP} proposed HDP to circumvent the severe limitation of CPDP: the need for a homogeneous set of metrics; i.e., different projects should have the same (or at least sufficient common) metrics to make CPDP work. The notion of transfer learning~\cite{Transferlerarning2010, Transferlearning2016} highlights the importance of exploiting all available resources, even if the input feature space and data distribution characteristics are not identical. In the original study~\cite{HDP}, authors examine the possibility of quickly transferring lessons learned from defect datasets to new, unseen datasets. 

The proposed methodology aims to employ the similarity between the distribution of heterogeneous metrics to address the restrictions on CPDP. After applying feature selection on the training set (i.e., source project), they use metrics matching to find the best-matched metrics in the testing set (i.e., target project) for each selected metric of the training set. This step may lead to infeasible HDP if the method is unable to find any paired metrics. They discussed the HDP target coverage to report the success rate of their method. After obtaining the best-matched set of metrics, they build a classifier on the source project and predict the defect-proneness of the target project. \citet{HDP} evaluate the performance of HDP based on the Area Under the ROC curve metric. They reported the representative HDP results using Gain Ratio for feature selection (top 15\% metrics), KS-test with the cutoff threshold of 5\% ($p=0.05$) for similarity finding, and the Logistic Regression (LR) as the classifier. Table~\ref{tab:paper2} gives a summary of the research questions and the steps followed in their paper. More details on their method are provided in Section \ref{sec:RQ4}.

\begin{center}
\begin{table*}[!htb]%
\centering
\caption{The overview of research questions and the steps taken in paper by~\citet{HDP}}%
\label{tab:paper2}
\resizebox{\textwidth}{!}{
\begin{tabular*}{530pt}{@{\extracolsep\fill}p{120pt}p{400pt}@{\extracolsep\fill}}
\midrule
\textbf{Project data used}     &   EQ, JDT, LC, ML, PDE, Apache, Safe, ZXing, ant-1.3, arc, camel-1.0, poi-1.5, redaktor, skarbonka, tomcat, velocity-1.4, xalan-2.4, xerces-1.2, cm1, mw1, pc1, pc3, pc4, jm1, pc2, pc5, mc1, mc2, kc3, ar1, ar3, ar4, ar5, ar6 \\ \hline
\textbf{Language}   & Java / Weka \\ \hline
\textbf{Preprocessing phase}   & 
    1) Feature Selection (Gain Ratio* , chi-square, relief-F, and significance attribute evaluation) \newline
    2) Training/Testing sets' metrics matching (Kolmogorov-Smirnov test based matching*, percentile based matching, and Spearman's correlation based matching)
    \\ \hline
\textbf{Classifiers}   & Simple logistic, Logistic regression*, Random Forest, Bayesian network, Support vector
machine, J48 decision tree, and Logistic model tree \\ \hline
\textbf{Cross-Validation Method}   & 500 times 2-fold stratified CV \\ \hline
\textbf{Evaluation Method}   & Area Under the Curve (AUC) of the Received Operating Characteristic (ROC) \\ \hline
\textbf{Statistical Tests/Methods}   & Wilcoxon signed-rank test, Cliff's $\delta$, and Kolmogorov-Smirnov test \\ \hline
\textbf{Research Questions}  & 
\textbf{RQ1: Is Heterogeneous Defect Prediction Comparable to WPDP, Existing CPDP Approaches for Heterogeneous Metric Sets, and Unsupervised Defect Prediction?}
\textbf{RQ2: What Are the Lower Bounds of the Size of Source and Target Datasets for Effective HDP?}  \\ 

\bottomrule
\end{tabular*}
}
\begin{tablenotes}
\singlespacing
\footnotesize
\item Items marked with an asterisk (*) are the main methods that are used in the research question. The rest are competitive approaches to check the validity of the model.  
\end{tablenotes}

\end{table*}
\end{center}

\section{INFORMATION ABOUT THE REPLICATION}\label{sec:Info_about_rep}
\subsection{Benchmark datasets}\label{datasets}
In this paper, we abide by the publicly available datasets that were used in earlier studies. The summary of the datasets is provided in Table~\ref{tab-data-summary}. Nonetheless, in some cases, e.g., NASA datasets, more than one version of the datasets were available and some concerns about the quality of the dataset have been reported~\cite{NASAQuality,MisuseNASA}. Hence, we established three important inclusion criteria to filter out unreliable datasets~\cite{Tantithamthavorn2017}:

\begin{enumerate}
    \item \textbf{Criterion 1- different corpora}: Since we are to implement HDP, we need to choose our datasets from a multitude of sources where there is a chance of inconsistency between column names. Furthermore, this factor will augment the generalizability of our conclusions.
    \item \textbf{Criterion 2- Sufficient EPV}: The number of Events Per Variable (EPV) is the ratio of buggy instances to the number of predictors (metrics). For instance, in JDT dataset, EPV is $206 / 19 =10.85$. To avoid problems related to the overfitting, underfitting, and misleading associations, general guidelines have been suggested for the minimum EPV required in multivariate analysis to be from 10 to 20~\cite{EPV-Tantithamthavorn}. We set the minimum acceptable EPV to 10 and discard any dataset whose EPV is less than 10 since few events relative to the number of independent variables produce unreliable results.
    \item \textbf{Criterion 3- defect ratio}: It is implausible to have more defective software modules than those that are free of bugs. Accordingly, we remove datasets whose defective rate is more than half ($>50\%$).
\end{enumerate}

Each group has a different set of measures in terms of the number and type, making CPDP infeasible in many cases. Therefore, we consider HDP as a remedy for incongruous metrics.

We analyse 136 publicly available defect datasets from various sources~\cite{AEEEM, Ambros2010, Kim2011, Jureczko2010, Jira2019, NASAQuality, MORPH, ReLink, ArcelikSOFTLAB}. We apply the inclusion criteria to sift qualified datasets out. 114 datasets have EPV less than 10, and 5 datasets have defective ratio greater than $50\%$; consequently, we arrive at 17 eligible datasets. In Table~\ref{tab-data-summary}, the number of instances, the number and the percentage of buggy instances per dataset, the number of metrics, and the EPV score of those datasets are reported.

\begin{center}
\begin{table*}[!htb]
\caption{Summary of the project data\label{tab-data-summary}}
\centering
\resizebox{\textwidth}{!}{
\begin{tabular}{cllrrrcr}
\toprule
\textbf{Group} & \textbf{Dataset} & \textbf{Abbreviation} & \textbf{buggy (\%)} & \textbf{\# of instances} & \textbf{buggy (\#)} & \textbf{\# of metrics} & \textbf{EPV} \\
\midrule
\multirow{5}{*}{Eclipse} & JDT~\cite{AEEEM,Ambros2010} & JDT & 20.7 & 997 & 206 & \multirow{3}{*}{19} & 10.8 \\
 & Mylyn~\cite{AEEEM,Ambros2010} & ML & 13.2 & 1862 & 245 &  & 12.9 \\
 & PDE~\cite{AEEEM,Ambros2010} & PDE & 14.0 & 1497 & 209 &  & 11.0 \\
 & Debug 3.4~\cite{Kim2011} & DG &24.7 & 1065 & 263 &  \multirow{2}{*}{17} & 14.6 \\
 & SWT 3.4~\cite{Kim2011} & SWT & 44.0 & 1485 & 653 & & 38.4 \\
\midrule
\multirow{5}{*}{Proprietary} & Prop-1~\cite{Jureczko2010} & PR1 & 14.8 & 18471 & 2738 & \multirow{5}{*}{20} & 130.4 \\
 & Prop-2~\cite{Jureczko2010} & PR2 & 10.6 & 23014 & 2431 &  & 115.8 \\
 & Prop-3~\cite{Jureczko2010} & PR3 & 11.5 & 10274 & 1180 &  & 56.2 \\
 & Prop-4~\cite{Jureczko2010} & PR4 & 9.6 & 8718 & 840 &  & 40.0 \\
 & Prop-5~\cite{Jureczko2010} & PR5 & 15.3 & 8516 & 1299 &  & 61.9 \\
\midrule
\multirow{3}{*}{Apache} & Camel 1.2~\cite{Jureczko2010} & CML & 35.5 & 608 & 216 & \multirow{3}{*}{20} & 10.8 \\
 & Xalan 2.5~\cite{Jureczko2010} & XN2.5 & 48.2 & 803 & 387 &  & 19.4 \\
 & Xalan 2.6~\cite{Jureczko2010} & XN2.6 & 46.4 & 885 & 411 &  & 20.6 \\
\midrule
\multirow{2}{*}{Jira} & Derby 10.2.1.6~\cite{Jira2019} & DY.2 & 33.7 & 1963 & 661 & \multirow{2}{*}{65} & 10.2 \\
 & Derby 10.3.1.4~\cite{Jira2019} & DY.3 & 30.3 & 2206 & 669 &  & 10.3 \\
\midrule
\multirow{2}{*}{NASA} & JM1~\cite{NASAQuality} & JM1 & 21.5 & 7782 & 1672 & 21 & 79.6 \\
 & PC5~\cite{NASAQuality} & PC5 & 27.5 & 1711 & 471 & 38 & 12.4 \\
 \bottomrule
\end{tabular}
}
\begin{tablenotes}
\singlespacing
\footnotesize
\item All the datasets can be downloaded via \href{github.com/HadiJahanshahi/Replication-HDP}{github.com/HadiJahanshahi/Replication-HDP}.
\end{tablenotes}
\end{table*}
\end{center}

\subsection{Level of interaction with authors of original study}\label{interaction}
In this replication study, the original researchers, \cite{HDP}, provided us most of the datasets; we queried about the detail of some parts that were not clearly mentioned in the paper, and the authors responded to our email. Since the scripts of the paper were not available, all the codes have been written from scratch, and we did not request the related codes from the authors.

\subsection{Changes to the original experiment}\label{changes}
We had original datasets, to some extent, to create our benchmark; however, several new datasets are incorporated to validate the generalizability of the original paper. The replicated research questions and the design of the experiments are similar to those of the original study with the following modifications:
\begin{itemize}
\item We define three new Research Questions to first investigate whether HDP has the transferability to be applied on a dataset with a totally different nature, and second whether the performance of the HDP can be improved through pooling and voting approaches. 
\item Different points of view into the problem for each Research Question are embodied in research.
\item The original study used only Logistic Regression (LR) as a classifier, whereas we implement both LR and Random Forest (RF) with the number of trees equal to 100.
\item Unlike the original study, we consider 136 different datasets, and based on the predefined criteria we choose only 17 reliable projects. Hence, only 5 datasets, namely JDT, Mylyn, PDE, JM1, and PC5, are common with the original study, and the remaining 12 projects are new.
\end{itemize}

From the research questions of this study reported in Section~\ref{sec:Intro}, RQ2 is identical to the original study, and the rest are newly defined.

\section{COMPARISON OF RESULTS WITH ORIGINAL STUDY}\label{sec:comparison}
Our partial replication study includes one of the research questions of the original study and three new research questions. Therefore, we extend their work and provide a comparison for RQ2 that is similar to the previous work. 

In the first step, we experiment with all cross-project model permutations for the 17 available datasets ($2\times {{17}\choose{2}} $ = 272 pairs). For each pair, one project is selected to be the testing set and the others as the training set. AUC values of the experiment have been computed and listed in Table~\ref{AUCQ0}. Note that many of the cells are empty in the table since we were unable to find enough common metrics to develop the CPDP method. In most cases, sufficient common metrics are found when we choose the training and testing set within the same group. To measure WPDP, we applied 10-fold cross-validation on a single dataset. The 10-fold cross-validation divides the dataset to 10 folds, and in each step, one fold will be used as the testing set and the remaining nine folds as the training set. Since cross-validation is highly affected by the data which is randomly selected in each fold~\cite{Cross-validation}, we repeated the process ten times to overcome the randomness issue. Accordingly, each boldface AUC value on the diagonal of Table~\ref{AUCQ0} is the result of aggregating (taking the average of) $10\times$10-fold cross-validation. Off-diagonal values indicate CPDP's performance using one dataset to predict the defects of another dataset. Therefore, no CV has been done for off-diagonal values, and they are obtained from a single experiment.

In the HDP context, there exists a prediction potential of other mature projects that remains intact. Any missing values in Table~\ref{AUCQ0} offer an unused potential that requires exploration. In Section~\ref{prediction_performance_HDP}, we aim to utilise the capability of the other projects in defect prediction.

\begin{center}
\begin{table*}[!htb]%
\footnotesize
\centering
\caption{Summary of AUC values for WPDP (boldface) and CPDP. Rows of the table indicate training projects, and the Columns are testing project. Each cluster of AUC performances demonstrates a unique system. (Rows are training, and columns are test sets.) \label{AUCQ0}}
\resizebox{\textwidth}{!}{
\begin{tabular}{l|lllllllllllllllll}
\toprule
 & \textbf{JDT} & \textbf{ML} & \textbf{PDE}& \textbf{CML} & \textbf{PR1} & \textbf{PR2} & \textbf{PR3} & \textbf{PR4} & \textbf{PR5} & \textbf{XN2.5} & \textbf{XN2.6} & \textbf{DY.2} & \textbf{DY.3} & \textbf{DG} & \textbf{SWT} & \textbf{JM1} & \textbf{PC5} \\
 \midrule
\textbf{JDT} & \textbf{0.81} & 0.86 & \multicolumn{1}{c|}{0.74} & &  &  &  &  &  &  &  &  &  &  &  &  &  \\
\textbf{ML} & 0.77 & \textbf{0.93} & \multicolumn{1}{c|}{0.71} &  &  &  &  &  &  &  &  &  &  &  &  &  &  \\
\textbf{PDE} & 0.81 & 0.81 & \multicolumn{1}{c|}{\textbf{0.77}} & &  &  &  &  &  &  &  &  &  &  &  &  &  \\
\cline{2-12}
\textbf{CML} &  &  & \multicolumn{1}{c|}{} &\textbf{0.65} & 0.55 & 0.58 & 0.59 & 0.51 & 0.59 & 0.62 & 0.66 & \multicolumn{1}{|c}{} &  &  &  &  &  \\
\textbf{PR1} &  &  & \multicolumn{1}{c|}{} & 0.54 & \textbf{0.75} & 0.65 & 0.64 & 0.71 & 0.65 & 0.50 & 0.54 & \multicolumn{1}{|c}{} &  &  &  &  &  \\
\textbf{PR2} &  &  & \multicolumn{1}{c|}{} & 0.56 & 0.71 & \textbf{0.72} & 0.69 & 0.70 & 0.69 & 0.58 & 0.69 & \multicolumn{1}{|c}{} &  &  &  &  &  \\
\textbf{PR3} &  &  & \multicolumn{1}{c|}{} & 0.58 & 0.67 & 0.69 & \textbf{0.72} & 0.65 & 0.71 & 0.58 & 0.66 & \multicolumn{1}{|c}{} &  &  &  &  &  \\
\textbf{PR4} &  &  & \multicolumn{1}{c|}{} & 0.54 & 0.70 & 0.62 & 0.62 & \textbf{0.76} & 0.62 & 0.61 & 0.68 & \multicolumn{1}{|c}{} &  &  &  &  &  \\
\textbf{PR5} &  &  & \multicolumn{1}{c|}{} & 0.59 & 0.68 & 0.69 & 0.70 & 0.65 & \textbf{0.71} & 0.61 & 0.69 & \multicolumn{1}{|c}{} &  &  &  &  &  \\
\textbf{XN2.5} &  &  & \multicolumn{1}{c|}{} & 0.59 & 0.52 & 0.50 & 0.53 & 0.59 & 0.52 & \textbf{0.69} & 0.70 & \multicolumn{1}{|c}{} &  &  &  &  &  \\
\textbf{XN2.6} &  &  & \multicolumn{1}{c|}{} & 0.60 & 0.58 & 0.63 & 0.63 & 0.60 & 0.64 & 0.64 & \textbf{0.82} & \multicolumn{1}{|c}{} &  &  &  &  &  \\ 
\cline{5-14}
\textbf{DY.2} &  &  &  &  &  &  &  &  &  &  &  & \multicolumn{1}{|c}{\textbf{0.86}} & 0.73 & \multicolumn{1}{|c}{} &  &  &  \\
\textbf{DY.3} &  &  &  &  &  &  &  &  &  &  &  & \multicolumn{1}{|c}{0.84} & \textbf{0.83} & \multicolumn{1}{|c}{} &  &  &  \\
\cline{13-16} 
\textbf{DG} &  &  &  &  &  &  &  &  &  &  &  &  &  & \multicolumn{1}{|c}{\textbf{0.73}} & \multicolumn{1}{c|}{0.66} & &  \\
\textbf{SWT} &  &  &  &  &  &  &  &  &  &  &  &  &  & \multicolumn{1}{|c}{0.65} & \multicolumn{1}{c|}{\textbf{0.94}} &  &  \\
\cline{15-18}
\textbf{JM1} &  &  &  &  &  &  &  &  &  &  &  &  &  &  & \multicolumn{1}{c|}{} & \textbf{0.69} & \multicolumn{1}{c|}{0.70} \\
\textbf{PC5} &  &  &  &  &  &  &  &  &  &  &  &  &  &  & \multicolumn{1}{c|}{} & 0.65 & \multicolumn{1}{c|}{\textbf{0.73}}  \\
\bottomrule
\end{tabular}
}
\end{table*}
\end{center}

\begin{figure}[!bth]
        \centerline{\includegraphics[width=\linewidth]{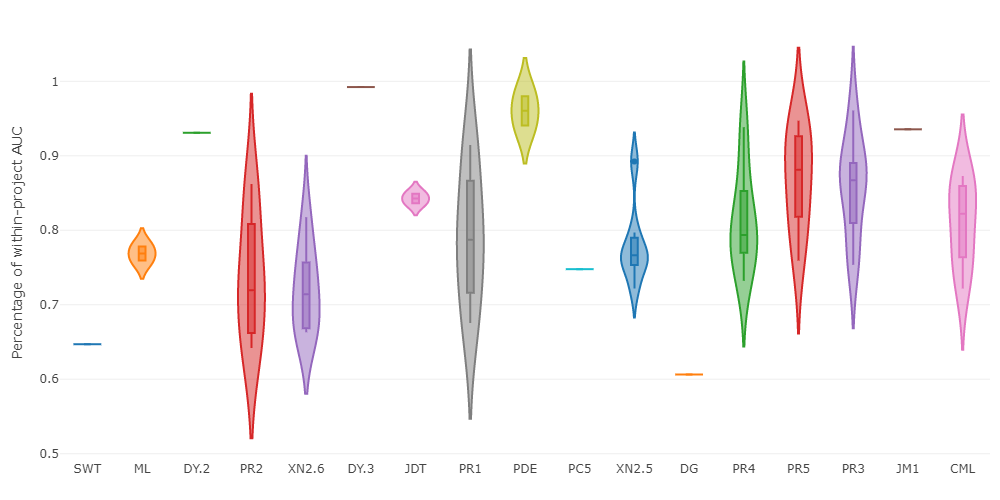}}
        \caption{Relation between the performance of CPDP and WPDP. After normalising off-diagonal AUC values in Table~\ref{AUCQ0} by Diagonal of the matrix (columnwise), we sorted projects based on their WP performance along the x-axis.\label{percentageAUC}}
\end{figure}

\vspace{-0.5cm}
Using beanplots, Figure~\ref{percentageAUC} demonstrates the cross-project performance of the models scaled (normalised) by the AUC values of the within-project model. The projects are sorted based on their within-project AUC, starting with the SWT 3.4 project (AUC value of 0.9) and ending to Camel 1.2 (AUC value of 0.65). Since they are sorted in descending order based on their WPDP's AUC values, we expect to see a decrease in the AUC of CPDP; however, such a pattern does not emerge. Similar to previous studies~\cite{JITKamei2016, Turhan-cross}, it can be concluded that the remarkable performance of WPDP in models is not a reliable indicator of that of CPDP.

\subsection{RQ1: How do models selected using domain-agnostic similarity perform in a cross-project context?}

To address the first Research Question, we used the domain-agnostic similarity of the projects introduced by \citet{JITKamei2016}. Algorithm~\ref{alg:domain-agnostic} lists the pseudocode to compute the domain-agnostic distance between two projects. First, the Spearman correlation of all metrics concerning the dependent variable (buggy or not) is calculated. Then, we choose the top 3 correlated metrics from the project and select the same ones from another project. Next, the pairwise Spearman correlation between selected metrics for each project is computed ($3\choose{2}$), and for each project, an array of 3 elements is generated. The Euclidean distance between these arrays informs us about the similarity of the projects.

\begin{algorithm}[!ht]
    \For{metrics $\in$ {Trainset}}{
       compute Spearman correlation wrt a dependent variable (label)
    }
    pick the top 3 metrics based on their Spearman correlation $\in$ {Trainset} $\land$ the same metrics $\in$ {Testset}
    
    \For{$i \in \{1,\hdots,{3 \choose 2}\}$}{
        $Q_i$ $\Leftarrow$ pairwise Spearman correlation of Trainset's picked metrics\\
        $R_i$ $\Leftarrow$ pairwise Spearman correlation of Test set's picked metrics
    }
    \textbf{return} EuclideanDistance $(Q,R)$
\caption{Domain-agnostic Dissimilarity calculation}
\label{alg:domain-agnostic}
\end{algorithm}

Using Algorithm~\ref{alg:domain-agnostic}, we report the pairwise similarity of all the projects in Table~\ref{Euc-Sim}. We select the most similar project to the testing project as our training set. For instance, in order to predict bugs in {\scshape JDT} project, we select {\scshape PDE} ($dist = 0.11$) which has the shortest distance from the test set. 

\begin{center}
\begin{table*}[!htb]%
\footnotesize
\centering
\caption{Domain-agnostic Distance of the Projects. (Rows are training and columns are testing sets.)\label{Euc-Sim}}
\resizebox{\textwidth}{!}{
\begin{tabular}{l|lllllllllllllllll}
\toprule
 & \textbf{JDT} & \textbf{ML} & \textbf{PDE}& \textbf{CML} & \textbf{PR1} & \textbf{PR2} & \textbf{PR3} & \textbf{PR4} & \textbf{PR5} & \textbf{XN2.5} & \textbf{XN2.6} & \textbf{DY.2} & \textbf{DY.3} & \textbf{DG} & \textbf{SWT} & \textbf{JM1} & \textbf{PC5} \\
 \midrule
\textbf{JDT} & 0.00 & 0.06 & 0.13 &  &  &  &  &  &  &  &  &  &  &  &  &  &  \\
\textbf{ML} & 0.49 & 0.00 & 0.54 &  &  &  &  &  &  &  &  &  &  &  &  &  &  \\
\textbf{PDE} & 0.11 & 0.04 & 0.00 &  &  &  &  &  &  &  &  &  &  &  &  &  &  \\
\textbf{CML} &  &  &  & 0.00 & 0.20 & 0.14 & 0.13 & 0.25 & 0.26 & 0.28 & 0.18 &  &  &  &  &  &  \\
\textbf{PR1} &  &  &  & 0.30 & 0.00 & 0.13 & 0.13 & 0.15 & 0.13 & 0.54 & 0.81 &  &  &  &  &  &  \\
\textbf{PR2} &  &  &  & 0.11 & 0.11 & 0.00 & 0.06 & 0.11 & 0.13 & 0.44 & 0.68 &  &  &  &  &  &  \\
\textbf{PR3} &  &  &  & 0.39 & 0.13 & 0.08 & 0.00 & 0.05 & 0.06 & 0.60 & 0.86 &  &  &  &  &  &  \\
\textbf{PR4} &  &  &  & 0.23 & 0.12 & 0.06 & 0.06 & 0.00 & 0.14 & 0.48 & 0.69 &  &  &  &  &  &  \\
\textbf{PR5} &  &  &  & 0.42 & 0.13 & 0.13 & 0.06 & 0.04 & 0.00 & 0.65 & 0.91 &  &  &  &  &  &  \\
\textbf{XN2.5} &  &  &  & 0.34 & 0.43 & 0.32 & 0.35 & 0.41 & 0.35 & 0.00 & 0.11 &  &  &  &  &  &  \\
\textbf{XN2.6} &  &  &  & 0.32 & 0.09 & 0.30 & 0.30 & 0.43 & 0.13 & 0.13 & 0.00 &  &  &  &  &  &  \\
\textbf{DY.2} &  &  &  &  &  &  &  &  &  &  &  & 0.00 & 0.04 &  &  &  &  \\
\textbf{DY.3} &  &  &  &  &  &  &  &  &  &  &  & 0.04 & 0.00 &  &  &  &  \\
\textbf{DG} &  &  &  &  &  &  &  &  &  &  &  &  &  & 0.00 & 0.40 &  &  \\
\textbf{SWT} &  &  &  &  &  &  &  &  &  &  &  &  &  & 0.57 & 0.00 &  &  \\
\textbf{JM1} &  &  &  &  &  &  &  &  &  &  &  &  &  &  &  & 0.00 & 0.05 \\
\textbf{PC5} &  &  &  &  &  &  &  &  &  &  &  &  &  &  &  & 0.02 & 0.00  \\
\bottomrule
\end{tabular}
}
\end{table*}
\end{center}

\begin{figure}[!htb]
        \centerline{\includegraphics[width=0.35\textwidth]{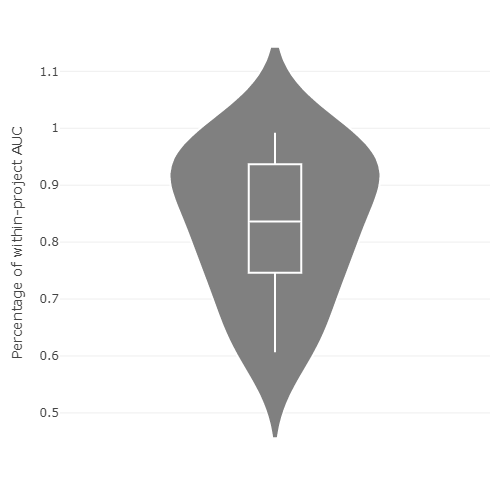}}
        \vspace{-0.5cm}
        \caption{RQ1: the performance of the models selected by the domain-agnostic similarity.\label{domain-agnostic}}
\end{figure}

\vspace{-0.5cm}
Figure~\ref{domain-agnostic} shows the AUC values of the models selected by the domain-agnostic similarity. The values are normalised by WP performance. We use a $t$-test to check the hypothesis whether the true mean of domain-agnostic models normalised by WPDP is equal to 1 or not. The \textit{p-value} of 0.83 indicates WPDP outperforms the suggested defect prediction model at the significance level of 0.05 ($\alpha=5\%$).

\vspace{-0.5cm}

\vspace{-0.5cm}
\subsection{RQ2: How do HDP methods predict the defect in the system compared to WPDP methods?} \label{sec:RQ4}

One of the main issues of CPDP in traditional defect prediction is its poor performance~\cite{Zimmermann-CPDP,Turhan-cross}. However, some methods, such as Peters filter and Transfer Naive Bayes, have been applied to improve the performance of CPDP~\cite{Peters-BetterCPDP}. The main drawback of cross-project defect prediction that is yet to be addressed is the heterogeneous metrics that different projects may have. In some cases, the overlap of the metrics between the projects is slight or absent, making the CPDP almost infeasible. Therefore, a prediction model cannot be built on either group to predict defects in the other. The heterogeneous metrics in different datasets and metrics collection tools have been reported elsewhere~\cite{HDP,Lincke2008,repo-problems2012}. 

\begin{figure}[!htb]
        \centerline{\includegraphics[width=0.5\textwidth]{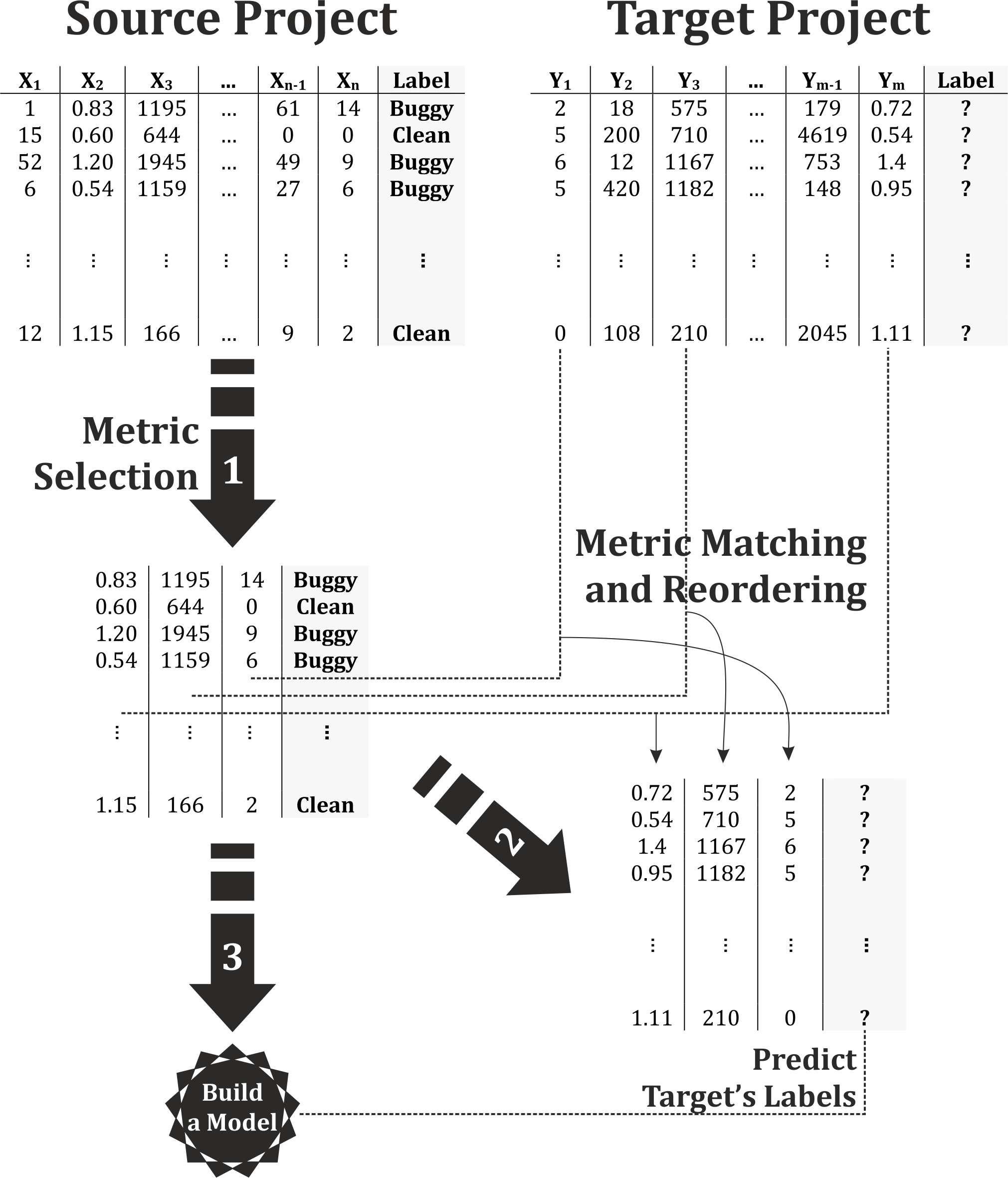}}
        \caption{Heterogeneous Defect Prediction (HDP)~\cite{HDP}\label{HDP-fig}}
\end{figure}

\citet{HDP} proposed an HDP approach to address the limitation of CPDP. First, they apply a feature selection technique on the source project (i.e., training set) to remove noise or irrelevant metrics. Then, based on the similarity between the distribution of the selected metrics and those of the target project (testing set), a list of the most similar metrics between the two datasets is selected. In this case, metrics' labels may differ. Finally, a classifier is built and trained using the set of the matched metrics and validated on the target project. Figure~\ref{HDP-fig} demonstrates the three main steps of the HDP approach. Note that both the labels and the number of the metrics may differ ($X_1 to X_n$ compared to $Y_1 to Y_m$). 

Although there is no best single feature selection technique for all defect prediction models~\cite{FeatureSelection2010, FeatureSelection2014}, \citet{HDP} reported that the Gain-Ratio metric selection has the best performance on HDP. In the feature selection part, the top 15\% of the metrics have been selected from the source project, as suggested by Gao et al.~\cite{featureSelection2011}. For each selected metric in the source project, the HDP approach computes the similarity between this metric and all metrics in the target project. In the original paper, the KS-test based matching approach has been implemented to find the best pair for each selected metric. Since the KS-test is non-parametric, no presumption of the distribution, e.g., normality, is needed. Statistically, the \textit{p-value} of the test indicates whether the distribution of the two metrics is similar. The more similar the metrics are, the closer to 1 the \textit{p-value} would be. Identical to the original experiment, we define the matching score of metric $i$ of the source project with metric $j$ of the target project as:
\begin{eqnarray}
M_{ij} = \textit{p-value}_{ij} & \text{of KS-test}.
\end{eqnarray}

\begin{figure*}[!htb]
        \centerline{\includegraphics[width=0.8\linewidth]{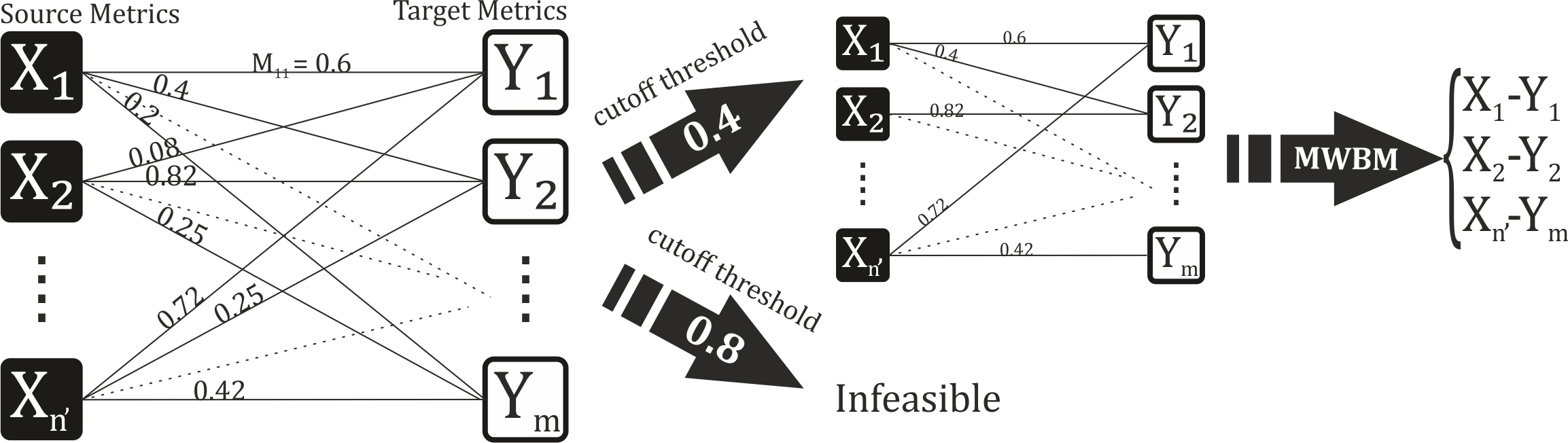}}
        \caption{How to match metrics of Source and Target Project. (MWBM: the Maximum Weighted Bipartite Matching) \label{MetricsSelection}}
\end{figure*}

After finding all pairwise similarities (\textit{p-values}) between the selected source metrics and target metrics, we define a cutoff threshold to eliminate loosely correlated metrics. In Figure~\ref{MetricsSelection}, if the cutoff threshold is 0.4, the poorly matched metrics will be removed, and we arrive at the top right graph. However, if our expectation of similarity is higher, the cutoff threshold of 0.8 reaches to infeasible HDP since the number of matched arcs (acceptable scores) is less than the number of source nodes (source metrics). We define Target Prediction Coverage (TPC) as the percentage of the target projects that can be predicted by HDP. The lower TPC is, the less viable HDP will be. In the original paper, the optimistic cutoff threshold of 0.05 is selected, and we use the same value here. Moreover, our HDP design is the same as that of the original study. 

After applying the cutoff threshold, if HDP is feasible (i.e. cutoff threshold of 0.4 in Figure~\ref{MetricsSelection}), the Maximum Weighted Bipartite Matching (MWBM) technique is used to decide on a group of matched metrics whose sum of matching scores is higher. In the top right graph of Figure~\ref{MetricsSelection}, for metrics $Y_m$ we have only one option: $X_{n^\prime}$. Hence, this pair will be assigned and the arc of $X_{n^\prime} \text{-} Y_m$ will be removed. Thus, for $Y_1$ there will be one option to be selected: $X_1$. Lastly, the pair of $X_2\text{-}Y_2$ will be created and the corresponding set of matches ($\{X_1 \text{-} Y_1 , X_2 \text{-} Y_2 , X_{n^\prime} \text{-} Y_m \}$) will be obtained.

After attaining a feasible set of metrics, the prediction model will be trained on the source project. We use both RF and the LR classifier to predict defects in the target set. We compare the result with WPDP as the baseline. Similar to the previous study, in WPDP, no feature selection technique has been applied to the training set.

\subsubsection{Experimental Design}{\label{sub_exp}}
From Table~\ref{tab-data-summary}, we use 17 eligible projects to validate HDP. WPDP can be applied if and only if a dataset is split into training and testing sets. For this purpose, we repeat randomly 2-fold cross-validation  (CV) 100 times. For each project, 200 testing sets are obtained. When conducting the 2-fold CV, we used a stratified CV in which the buggy rate of both folds will be the same as that of the original datasets. In WPDP, we use the remaining 200 training sets to train the model; therefore, we obtain 3400 ($= 17 \times 200$) AUC values for the within-project experiment. On the other hand, in HDP, for each 200 test set, we train on all other 16 projects to check the feasibility of HDP and the accuracy of the model. Although in the optimal case, we should obtain 54400 ($=17 \times 16 \times 200$) AUC values for HDP, we face several infeasible cases in which, based on the cutoff threshold, no similar metric can be selected between the source and target project. In Table~\ref{tab:AUC-HDP}, a typical output of HDP is presented. In the provided result, ``Project 3'' is unable to have any prediction on ``Project 1'' since all 200 ($=2\times100$) values in the subsequent columns are ``NaN''. In addition to infeasible HDP, in the first row, ``Project 2'' is able to achieve an absolute prediction on ``Project 1''. Nevertheless, in most pairwise predictions, neither is the case. Out of 200 repetitions, some produce feasible HDP while others fail. Therefore, a threshold on the acceptable number of predictions is required to claim whether HDP between two projects is feasible.

\begin{center}
\begin{table*}[!htb]%
\centering
\caption{A sample result of the output of HDP. NaNs indicate infeasible HDP. It includes the result of 2-fold cross-validation that is randomly repeated 100 times for each pair of projects. Therefore, the total number of columns is 202 (=$2+100\times2$), and the total number of rows is 272 ($=17\times16$). \label{tab:AUC-HDP}}%
\resizebox{\textwidth}{!}{
\begin{tabular}{|c|c|c|c|c|c|c|c|c|c|c|}
\hline
\textbf{Train}                                         & \textbf{Test} & \textbf{CV1-1} & \textbf{CV1-2} & \textbf{CV2-1} & \textbf{CV2-2} & \multicolumn{2}{c|}{$\cdots$}         & \textbf{CV199-2} & \textbf{CV200-1} & \textbf{CV200-2} \\ \hline
Project 2                                              & Project 1     & 0.75           & 0.92           & 0.98           & 0.84           & \multicolumn{2}{c|}{\multirow{3}{*}{$\ddots$}} & 0.57             & 0.90              & 0.65             \\ \cline{1-6} \cline{9-11} 
Project 3                                              & Project 1     & NaN            & NaN            & NaN            & NaN            & \multicolumn{2}{c|}{}                   & NaN              & NaN              & NaN              \\ \cline{1-6} \cline{9-11} 
Project 4                                              & Project 1     & 0.72           & NaN            & 0.65           & 0.92           & \multicolumn{2}{c|}{}                   & 0.67             & NaN              & 0.89             \\ \hline
$\vdots$                                                      & $\vdots$             & $\vdots$              & $\vdots$              & $\vdots$              & $\vdots$              & \multicolumn{2}{c|}{$\vdots$}                  & $\vdots$                & $\vdots$                & $\vdots$                \\ \hline
\begin{tabular}[c]{@{}c@{}}Project 14\end{tabular} & Project 17    & 0.82           & 0.62           & NaN            & 0.79           & \multicolumn{2}{c|}{\multirow{2}{*}{$\ddots$}} & 0.91             & NaN              & 1.00              \\ \cline{1-6} \cline{9-11} 
\begin{tabular}[c]{@{}c@{}}Project 15\end{tabular} & Project 17    & 1.00              & 0.71           & 0.32           & NaN            & \multicolumn{2}{c|}{}                   & NaN              & 0.65             & 0.95             \\ \hline
\begin{tabular}[c]{@{}c@{}}Project 16\end{tabular} & Project 17    & NaN            & 0.51           & 0.73           & 0.48           & \multicolumn{2}{c|}{$\cdots$}                  & 0.82             & 0.62             & NaN              \\ 
\hline
\end{tabular}
}
\end{table*}
\end{center}

\begin{figure}[!htb]
        \centerline{\includegraphics[width=0.6\linewidth]{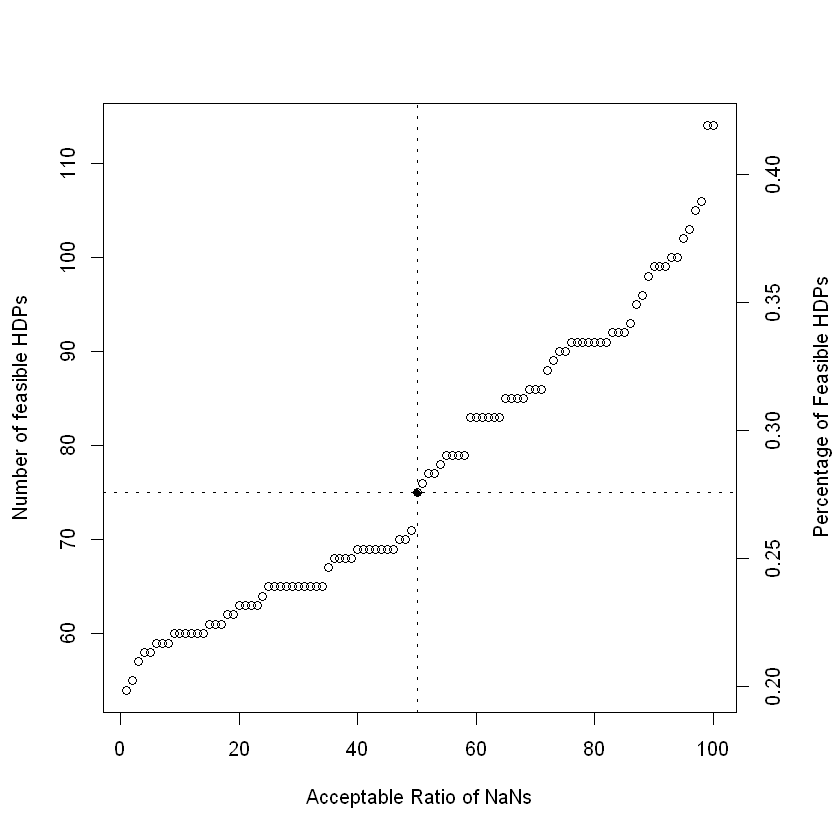}}
        \caption{Feasibility of HDP on 17 defect prediction datasets. The number of feasible HDP out of 272 pairwise HDPs, together with its acceptable ratio of NaNs, is reported. \label{FeasibleHDP}}
\end{figure}

\vspace{-0.5cm}
Figure~\ref{FeasibleHDP} shows different cutoff thresholds for the percentage of acceptable NaNs. When we permit only 1\% NaNs out of 200 replications, the feasibility of HDP reduces to 19.9\% ($=54$ feasible HDPs out 272). When we increase the threshold to 50\%, the power of HDP increases to 27.8\% ($=75$ feasible HDPs out 272), and finally, when we accept 99\% of NaNs that seems optimistic, only 114 feasible cases will be obtained ($41.9\%$). Therefore, the KS-test with a cutoff of 0.05 reaches a few feasible HDPs ranging between 19.9\% (in the pessimistic case) to 41.9\% (in the optimistic case) of the total replications. Notwithstanding the limitation of HDP in prediction, the original paper does not mention the cutoff threshold for the acceptable number of NaNs. The authors of the original study only mentioned the total number of feasible cases being 284 out of 962, which is close to the optimistic result that we reported in our replication experiment. 

\subsubsection{Prediction Performance}\label{prediction_performance_HDP}
We investigate the representative HDP results based on Gain Ratio feature selection, K-S Test with a cutoff threshold of 0.05, the LR classifier (in addition to RF), and Area Under the ROC Curve (AUC). We also set the percentage of acceptable NaNs in the prediction to 99\%. Since each dataset has a different set of metrics, the only acceptable baseline to compare with is the WPDP. 

Table~\ref{tab:RQ4:results} compares the performance of HDP (mean AUC) with the baseline. The first column demonstrates the result of WPDP based on the mean of AUC values for 200 replications on the target (test) dataset. The performance of HDP on the target set using other datasets is shown in the third column. The values are obtained through feasible HDPs. For example, {\scshape JDT} project can be predicted by other projects (from the same or different groups) in 34\% of the cases. The infeasible HDPs acquired from the remaining projects are ignored. For the target project {\scshape JDT}, there exist 16 other projects as the source project (train set). Furthermore, we split the target to 2 folds 100 times, and, in total, the potential 3200 cases of pairwise prediction can be generated. The fourth column indicates that out of these 3200 cases, only 1087 of them are feasible for {\scshape JDT} project. Therefore, the average rate of predictability is only 28.1\%. It indicates that in 71.9\% of the replications, no common metrics (similar metrics based on KS-test with a cutoff of 0.05) have been found. The second column reports the magnitude of the effect size between WPDP and HDP in terms of Cliff's $\delta$~\cite{cliffdelta1993}. Cliff's $\delta$ ranges from +1 to -1. The more the Cliff's $\delta$ is, the better the HDP performance will be in comparison to WPDP. The estimate of Cliff's $\delta$ magnitude (N: Negligible, S: Small, M: Medium, and L: Large) illustrates how significant the statistics are. In all cases, we encounter significant, negative numbers, indicating the baseline outperforms the proposed HDP. 

As previously reported in Section~\ref{sub_exp}, by setting the cutoff threshold to 99\% for NaNs acceptance, 114 feasible HDP models would be achieved. To investigate further the results of the mean and Cliff's $\delta$ comparison, we applied Wilcoxon signed-rank test~\cite{Wilcoxon1945} to the results. Using an alpha level of 5\% ($\alpha=0.05$), we build a pairwise Wilcoxon test between all feasible HDPs and corresponding WPDPs. Using LR Classifier, in all cases, WPDP outperforms while this ratio is even higher when compared with RF. This observation is different than the original study as we apply different filtering mechanisms and report infeasible cases that are ignored in that paper.

\begin{center}
\begin{table*}
\centering
\caption{Comparing the performance of HDP and WPDP (by mean AUC of all feasible HDPs). \label{tab:RQ4:results}}%
\resizebox{\textwidth}{!}{
\begin{tabular}{l|rrrrrccc|ccc}
\toprule
\multicolumn{1}{c|}{}  & \multicolumn{8}{c|}{Logistic classifier (RWeka package in R)}  & \multicolumn{3}{c}{Random Forest}          \\ \hline
\multicolumn{1}{c|}{\textbf{Target}} & \textbf{WPDP (mean)} & \textbf{WPDP (Cliff's $\delta$)} & \textbf{HDP (mean)} & \textbf{Predictability \#} & \textbf{Predictability \%} & \textbf{Win} & \textbf{Tie} & \textbf{Loss} & \textbf{Win} & \textbf{Tie} & \textbf{Loss} \\ \hline
JDT & 0.821 & -0.992 (L) & 0.589 & 1087 & 34.0\% & 0 & 1 & 8             & 0 & 1 & 8 \\
ML & 0.920 & -1.000 (L) & 0.547 & 1245 & 38.9\% & 0 & 0 & 8             & 0 & 0 & 8 \\
PDE & 0.768 & -0.997 (L) & 0.556 & 1107 & 34.6\% & 0 & 2 & 6             & 0 & 2 & 6 \\
DG & 0.722 & -0.967 (L) & 0.552 & 1290 & 40.3\% & 0 & 0 & 7             & 0 & 0 & 7 \\
SWT & 0.879 & -1.000 (L) & 0.529 & 240 & 7.5\% & 0 & 0 & 3             & 0 & 0 & 3 \\
PR1 & 0.745 & -1.000 (L) & 0.500 & 401 & 12.5\% & 0 & 1 & 2             & 0 & 1 & 2 \\
PR2 & 0.711 & -1.000 (L) & 0.509 & 800 & 25\% & 0 & 0 & 1            & 0 & 0 & 4 \\
PR3 & 0.692 & -1.000 (L) & 0.509 & 1167 & 36.5\% & 0 & 0 & 4             & 0 & 0 & 7 \\
PR4 & 0.738 & -1.000 (L) & 0.499 & 587 & 18.3\% & 0 & 0 & 3           &  0 & 0 & 5 \\
PR5 & 0.703 & -1.000 (L) & 0.491 & 939 & 29.3\% & 0 & 0 & 3             & 0 & 0 & 6 \\
CML & 0.629 & -0.878 (L) & 0.546 & 1382 & 43.2\% & 0 & 2 & 10           & 0 & 1 & 11 \\
XN2.5 & 0.658 & -0.969 (L) & 0.560 & 1074 & 33.6\% & 0 & 2 & 8           & 0 & 2 & 8 \\
XN2.6 & 0.793 & -0.965 (L) & 0.596 & 1116 & 34.9\% & 0 & 2 & 7          & 0 & 2 & 7 \\
DY.2 & 0.841 & -0.997 (L) & 0.594 & 2246 & 70.2\% & 0 & 1 & 13          & 0 & 1 & 15 \\
DY.3 & 0.813 & -1.000 (L) & 0.571 & 2071 & 64.7\% & 0 & 0 & 11          & 0 & 0 & 13 \\
JM1 & 0.672 & -1.000 (L) & 0.566 & 270 & 8.4\% & 0 & 0 & 2             & 0 & 0 & 2 \\
PC5 & 0.733 & -0.999 (L) & 0.495 & 793 & 24.8\% & 0 & 1 & 6             & 0 & 1 & 6 \\
 \hline
Total  & 0.755 & - & 0.541 & - & 32.7\%  & \begin{tabular}[c]{@{}c@{}}0\\ (0\%)\end{tabular} & \begin{tabular}[c]{@{}c@{}}12\\ (9.3\%)\end{tabular} & \begin{tabular}[c]{@{}c@{}}117\\ (90.7\%)\end{tabular} & \begin{tabular}[c]{@{}c@{}}0\\ (0\%)\end{tabular} & \begin{tabular}[c]{@{}c@{}}11\\ (8.6\%)\end{tabular} & \begin{tabular}[c]{@{}c@{}}118\\ (91.4\%)\end{tabular} \\
\bottomrule
\end{tabular}
}
\end{table*}
\end{center}

\vspace{-0.5cm}
Statistically, it can be concluded that HDP is not a viable alternative for WPDP. For example, in Table~\ref{tab:RQ4:results}, PDE was predicted in 8 prediction combinations, and WPDP outperforms in six combinations, and they tie in two cases.

\subsection{RQ3: How do the ensembles of models built from several projects perform in an HDP context?}
In RQ2, we found that the performance of HDP is considerably worse than the baseline, WPDP. Other papers reported the same issue of poor performance in HDP due to class imbalance problem, complex nonlinear relationship between source and target datasets, and information loss in constructing similar distributions~\cite{Tong2019, Li2018}. Therefore, we propose an ensemble of learners in case of a lack of common metrics to have a better HDP. 

For this research question, we follow the same steps of HDP, shown in Figure~\ref{HDP-fig}. For each project, we first choose only projects that are considered as feasible based on the KS-test with a cutoff threshold of 0.05. Then, we build a model on each of them and predict the label of each instance in the target project. Afterward, each model gives us a probability of being buggy, and we take the average of the vote of all feasible models as the probability of defectiveness. In this case, instead of pairwise prediction, we have an ensemble of models voting for the correct label of the target set. Unlike RQ2, we do not use any cross-validation on the target set; instead, the whole target project is considered as the test set. 

\begin{table*}[!htb]
\centering
\caption{Comparison of the performance of ensemble voting approach using Logistic Regression (LR) and Random Forest (RF) in terms of the Area Under the ROC Curve (AUC). \label{tab:RQ3:comparison}}%
\resizebox{\textwidth}{!}{
\begin{tabular}[t]{@{}l|lrrrr}
\toprule
\textbf{Target Project} & \textbf{Training Project(s)} & \begin{tabular}[c]{@{}r@{}}\textbf{Average AUC}\\ \textbf{using ensemble}\\   \textbf{voting (LR)}\end{tabular} & \begin{tabular}[c]{@{}r@{}}\textbf{Average AUC}\\ \textbf{(LR)}\end{tabular} & \begin{tabular}[c]{@{}r@{}}\textbf{Average AUC} \\ \textbf{using ensemble}\\   \textbf{voting (RF)}\end{tabular} & 
\begin{tabular}[c]{@{}r@{}}\textbf{Average AUC}\\ \textbf{(RF)}\end{tabular} \\
\midrule
JDT & ML, PDE, PR2, PR4 & 60.3\% & 53.7\% & 60.3\% & 54.4\% \\
ML & JDT, PDE, CML, PR2, PR4 & 60.6\% & 54.0\% & 59.7\% & 53.8\% \\
PDE & JDT, ML,  CML, PR2, PR4 & 66.1\% & 54.4\% & 65.7\% & 56.1\% \\
CML & XN2.5, XN2.6 & 59.3\% & 59.1\% & 57.6\% & 56.7\% \\
PR1  &  ML, CML & 49.9\% & 49.9\% & 49.3\% & 50.0\% \\
PR2  &  JDT, ML, PDE, PR4& 51.4\% & 50.9\% & 51.5\% & 51.1\% \\
PR3 & JDT, ML, PDE, PR2, PR4, DY.3 & 51.4\% & 50.8\% & 51.0\% & 50.8\% \\
PR4  &  CML, PR2 & 49.4\% & 49.7\% & 50.1\% & 50.0\% \\
PR5  &  JDT, ML, PR2, PR4 & 48.9\% & 49.5\% & 48.9\% & 49.4\% \\
XN2.5  &  CML, XN2.6, DY.3 & 61.4\% & 57.8\% & 64.9\% & 58.9\% \\
XN2.6  &  CML, XN2.5 & 75.1\% & 64.8\% & 73.2\% & 63.2\% \\
\addvbuffer[0.0cm 0.23cm] {DY.2} & \begin{tabular}[c]{@{}l@{}} JDT, ML, PDE, PR2, \\ PR3, PR4, DY.3, JM1, PC5\end{tabular} & \addvbuffer[0.0cm 0.23cm] {73.3\%} & \addvbuffer[0.0cm 0.23cm] {59.0\%} & \addvbuffer[0.0cm 0.23cm] {77.8\% }& \addvbuffer[0.0cm 0.23cm] {56.6\%} \\
\addvbuffer[0.0cm 0.23cm] {DY.3} & \begin{tabular}[c]{@{}l@{}} JDT, ML, PDE, PR2, \\ PR3, PR4, DY.2, JM1, PC5 \end{tabular} & \addvbuffer[0.0cm 0.23cm] {71.0\%} & \addvbuffer[0.0cm 0.23cm] {58.0\%} & \addvbuffer[0.0cm 0.23cm] {71.2\%} & \addvbuffer[0.0cm 0.23cm] {55.0\%} \\
DG  &  CML, PR2, PR4, XN2.6, DY.3 & 68.6\% & 54.2\% & 56.5\% & 51.9\% \\
SWT  &  CML & 54.0\% & 54.0\% & 53.9\% & 53.9\% \\
JM1  &  CML & 56.6\% & 56.6\% & 55.0\% & 55.0\% \\
PC5  &  ML, CML & 47.1\% & 47.1\% & 47.0\% & 47.0\% \\
\bottomrule
\multicolumn{2}{c}{\textbf{Mean peformance}} & 59.1\% & 54.3\% & 58.4\% & 53.8\% \\
\bottomrule
\end{tabular}
}

\end{table*}

Table~\ref{tab:RQ3:comparison} shows the difference in the performance of ensemble voting compared to normal HDP. We compare the mean of both methods using the Wilcoxon signed-rank test with continuity correction, and the result for both classifiers, LR and RF, demonstrates significant improvement, with \textit{p-value} of $0.0058$ and $0.0069$ respectively ($\alpha = 0.05$). Hence, we recommend using the potential of all feasible datasets instead of pairwise HDP implementation. 

\subsection{RQ4: How Feasible is HDP in terms of target prediction coverage?}
HDP applicability is conditional on target prediction coverage --- the percentage of target projects that can be predicted by HDP models. If no feasible HDP exists, due to missing matched metrics, it might be impossible to utilise heterogeneous predictors. 

Table~\ref{tab:targetcoverage} shows how frequently a source dataset (rows of the table) can predict a target dataset (columns of the table) using the HDP algorithm. For example, Eclipse dataset can predict Apache dataset in 9 combinations, which is equal to 60\% of all available combinations between these two sets ($=N_{\text{Eclipse}} \times N_{\text{Apache}}=5\times3$). This pairwise prediction coverage reports many infeasible cases in our experiment.

\begin{center}
\begin{table*}[!htb]
\centering
\caption{Target Prediction Coverage of HDP. \label{tab:targetcoverage}}%
\resizebox{\textwidth}{!}{
\begin{tabular}{c|ccccc|ccc}
\toprule
\textbf{Source}  & \textbf{Eclipse}   & \textbf{Proprietary}      & \textbf{Apache}        & \textbf{Jira}         & \textbf{NASA}   & \textbf{\begin{tabular}[c]{@{}c@{}}WPDP\\AUC\end{tabular}} & \textbf{\begin{tabular}[c]{@{}c@{}}HDP\\AUC\end{tabular}} & \textbf{\begin{tabular}[c]{@{}c@{}}HDP \\ Target \\ Coverage\end{tabular}} \\ \hline
\textbf{Eclipse}   & \textbf{8 (40\%)} & 13 (52\%)             & 9 (60\%)             & 5 (50\%)               & 0 (0\%)           & 0.82         & 0.56        & 62.5\%                                                                     \\
\textbf{Proprietary}  & 1 (4\%)       & \textbf{7 (35\%)} & 4 (26.7\%)             & 2 (20\%)           & 0 (0\%)        & 0.71         & 0.50        & 93.8\%                                                                     \\
\textbf{Apache}   & 10 (67\%)          & 9 (60\%)             & \textbf{6 (100\%)} & 3 (50\%)           & 3 (50\%)           & 0.66         & 0.56        & 93.8\%                                                                       \\
\textbf{Jira}    & 4 (40\%)          & 9 (90\%)          & 6 (100\%)          & \textbf{2 (100\%)} & 4 (100\%)       & 0.83         & 0.63        & 75.0\%                                                                     \\
\textbf{NASA} & 4 (40\%)         & 1 (10\%)             & 2 (33\%)             & 0 (0\%)          & \textbf{2 (100\%)} & 0.70         & 0.48        & 52.9\%   \\ 
\bottomrule
\end{tabular}
}
\end{table*}
\end{center}

\vspace{-0.5cm}
Another important implication is the bold-faced values on the main diagonal of Table~\ref{tab:targetcoverage}. Those values refer to the intragroup prediction feasibility of HDP. Since within each group of projects (e.g., Eclipse, Jira, etc.), almost all the metrics are the same, CPDP on the main diagonal of Table~\ref{tab:targetcoverage} is also feasible. However, HDP is unable to find similar metrics in some projects after applying feature selection. For example, Eclipse can predict itself in 8 out of 20 possible combinations ($=N_{\text{Eclipse}} \times (N_{\text{Eclipse}}-1)=5\times4$). The poor performance of metric selection/matching might lead to a lower self-coverage. We use the diagonal value and their available WPDP to validate the feasibility of HDP.

In Table~\ref{tab:targetcoverage}, also, the median of AUC values (using either WPDP or HDP) for each group has been mentioned. WPDP statically performs better than HDP. The last column of Table~\ref{tab:targetcoverage} shows HDP target coverage of each source group. As the definition offers, a source group can cover a dataset of a different group if and only if at least a dataset in the source group can be used to build an HDP model and be tested on the target dataset. For example, NASA has coverage of 52.9\% indicating that as the source group, its datasets (JM1 and PC5) can predict 9 out of 17 datasets, namely four projects in Eclipse group, one project in Proprietary, two projects in the Apache group, and two projects from NASA group. Proprietary and Apache by 93.8\% coverage are the most reliable dataset groups on which we can build an HDP.

\begin{figure}[!htb]
        \centerline{\includegraphics[width=0.6\linewidth]{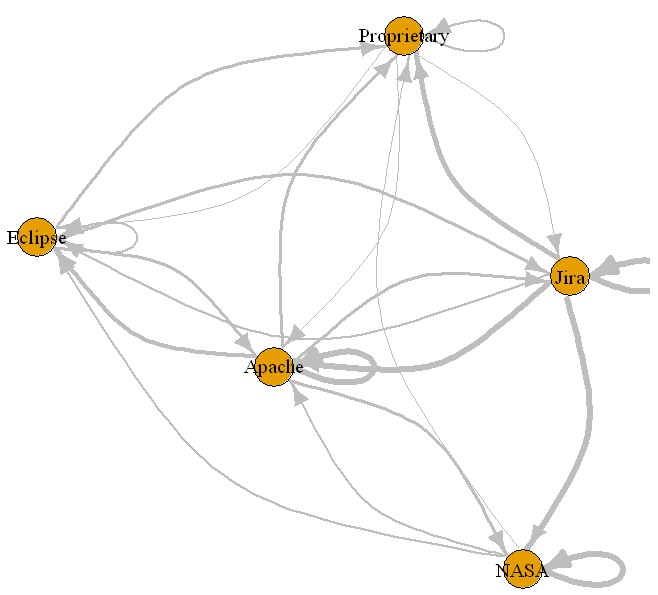}}
        \caption{Pairwise target prediction coverage of HDP \label{fig:HDPCoverage}}
\end{figure}

Figure~\ref{fig:HDPCoverage} illustrates the pairwise target prediction of group datasets. The thicker the arc is, the better the coverage will be. The outgoing arcs indicate the strength of target coverage, whereas the incoming arcs show how coverable a group is. For example, even though Jira can cover most of the datasets inside of the NASA group, NASA does not have such a mutual success rate. Accordingly, the upshot of the experiment indicates that there is a long path to build a real successful HDP.

\section{THREATS TO VALIDITY}\label{sec:validity}
Both stratified cross-validation and normal cross-validation are not good representatives of the chronology of a dataset. In practice, the training set is always older than the testing set, whereas when we shuffle the dataset to implement CV, this order will be discarded. Therefore, other validation techniques that capture the time order in the dataset may yield different outcomes. Furthermore, we filter out irrelevant and unreliable projects based on predefined assumptions; however, there might exist some different criteria that influence the quality of the defect datasets. Thus, the more comprehensive project selection techniques may lead to better performance of the methods. 

Here, we only use publicly available datasets while industry-related datasets may yield to a different conclusion; therefore, our partial replication study can reflect the result of the experiments where there is an overlap in our datasets. Such a difference does not affect the validity of the study, while it may overlook some conclusions in other types of projects.

The original work is implemented in Weka and Java, whereas we applied HDP using R (RWeka). The difference in the results may arise from the difference in their default variable options. Moreover, the correct threshold setting for HDP is a challenging issue that remains as future work. Finally, the granularity of the datasets differs. Considering this phenomenon might result in different outcomes. In future works, the datasets can be clustered based on their granularity, and we need to check whether HDP is sensitive to similarity in/difference between granularity of the source and target projects.

\section{CONCLUSIONS ACROSS STUDIES}\label{sec:Conclusion}
In this paper, we explored different approaches to build and validate HDP models. Defect prediction plays a crucial role in terms of efficiently allocating limited test resources. Traditional defect models use code metrics (e.g., Halstead and McCabe Metrics\cite{Halstead}) to classify a module as buggy or clean. Since there exist many potential defect datasets with heterogeneous metrics, we applied HDP on publicly available traditional defect datasets to learn whether transfer learning within traditional models is feasible. 

We have conducted an empirical validation on HDP application and applied guidelines for reporting an experimental replication~\cite{replication} to draw a cross-study conclusion between the original study and our replication. The findings of this partial, empirical replication study are as follows.

\begin{itemize}
    \item The Within-Project performance of models is not a significant factor while applying Cross-Project Models.
    \item Even though applying a domain-agnostic similarity to predefine the best-performing model brings better performance than the median AUC of CPDP, WPDP still outperforms the models selected by domain-agnostic similarity (RQ1).

    \item Before using HDP, we need to adjust several hyper-parameters. The feature selection threshold (15\%), cutoff threshold of KS-test \textit{p-value} ($p = 0.05$), and the maximum percentage of acceptable infeasible cases (NaNs) (99\%) raise the likelihood of unsuccessful HDP (RQ2). Hence, there is a need to optimise these techniques and balance the trade-off between the sensitivity of the algorithm and the selection of the parameters. The original study had not reported this issue.
    \item We proposed an ensemble voting approach in the HDP context using multiple source projects where there is a lack of common metrics. In terms of AUC, the performance of the revised HDP has increased by 4.8 percent (RQ3) compared to the original study.
    \item HDP Target Coverage was not promising. Besides, the performance of WPDP is statistically more significant than that of HDP (RQ2 and RQ4). These two issues indicate the need for further analyses to improve and generalise the heterogeneous prediction model. This comparison had not been performed in the original work. 
    \item HDP cannot transfer the knowledge learned from a source dataset due to the substantial difference between the distribution of the metrics in a target and source project. In future research, employing exploratory analyses and conducting preprocessing techniques are highly recommended before applying HDP approaches.
\end{itemize}


\section*{Supporting information}
To make the study reproducible, the datasets, codes, and outputs are publicly available at \href{github.com/HadiJahanshahi/Replication-HDP}{github.com/HadiJahanshahi/Replication-HDP}.


\bibliographystyle{ACM-Reference-Format}
\bibliography{bib}

\end{document}